\documentclass[a4paper,11pt]{article}
\pdfoutput=1 

\usepackage{jheppub} 

\usepackage[T1]{fontenc} 

\newcommand{\be}{\begin{equation}}
\newcommand{\ee}{\end{equation}}

\title{\boldmath Dynamical Quarkonia Suppression in a QGP-Brick}

\author{Jorge Casalderrey-Solana}

\affiliation{
Departament d'Estructura i Constituents
de la Mat\`eria and Institut de Ci\`encies del Cosmos (ICCUB),
Universitat de Barcelona, Mart\'\i \ i Franqu\`es 1, 08028 Barcelona, Spain
}

\emailAdd{jorge.casalderrey@ub.edu}

\abstract{
I address the effect that a temperature dependent potential has on the suppression of heavy 
quarkonia states in deconfined hadronic matter.  
I focus on a simple medium: a homogenous, fixed temperature and deconfined system with a finite lifetime
(QGP-brick).
Assuming that all the interactions of 
a heavy quark anti-quark ($Q-\bar Q$) pair  with the medium can be recast into an in-medium potential,
I solve the time dependent Schr\"odinger equation to evolve the density matrix which describes
the hard pair production and its connection to the final distribution of hadrons after the medium disappears.
For those temperatures in which bound states survive in the medium, I find a non-trivial dependence of
the production of excited quarkonia states on the in-medium levels, due to the mixing of vacuum and in-medium
wave functions. This mixing leads, in particular, to the enhancement of the relative abundance of 2S to 1S states
for those systems in which the in-medium ground state of the $Q-\bar Q$ system is dissolved or close to threshold.
I also explore quarkonia production in a non-homogeneous expanding medium and find that the finite formation time 
effects arising from the low binding energies of in-medium states lead to the insensitivity of the heavy mesons yield to the
hottest part of the system evolution.
}

\begin{document} 
\maketitle
\flushbottom

\section{Introduction}
\label{sec:intro}
At  temperatures of about $T_c\approx 175$ MeV, QCD experiences a rapid cross over transition into
a deconfined state of matter, the Quark Gluon Plasma (QGP).   While at asymptotically high 
temperatures the QGP behaves as a free gas of quarks and gluons, in the vicinity of $T_c$  strong
interactions among the colored plasma constituents remain, which complicate the description of the
medium dynamics. These interactions make the QGP an extremely rich system, with strong collective phenomena
which arise directly from an non-abelian fundamental gauge theory.  Therefore, the characterization of this region of the QCD phase diagram has been
the subject of intense theoretical and experimental research.

From the experimental point of view, 
this state of matter is studied by the ultra-relativistic collision of heavy ions which form  a hot and dense hadronic medium. At the collision energies explored at RHIC and the LHC, the energy density of this medium is sufficiently large to reach the deconfined phase at the early stage of the system evolution \cite{Adcox:2004mh,Adams:2005dq,Muller:2012zq}.
One of the main tools available for determining the medium dynamics
is its interactions with hard probes. 
Among those, an important set  is the different heavy quarkonia states and their in-medium modifications \cite{Matsui:1986dk}.
The advantage of these probes is that the heavier their  constituents quarks are, the smaller the distances they probe so that 
 their properties can be described within a peturbative QCD framework and, thus, are under good theoretical
control. The interaction of these states with the medium leads to a modification of those properties and even to their
complete dissociation at sufficiently high temperatures, which, in turn, leads to a suppression of quarkonium yields
in nuclear collisions as compared to their p-p counterpart. 

Traditionally, due to the small b-quark production cross section at low collision energies, most studies have focussed on charmonia
states (such as the J$/\psi$). However, charm quarks are not sufficiently heavy for a reliable theoretical description, which leads to significant 
uncertainties both in their production mechanism in nuclear collisions \cite{Lansberg:2006dh}  as well as their interactions with deconfined matter (see, \cite{Rapp:2008tf} for a recent review),  what
complicates data interpretations. 
In addition, the abundance of $c-\bar c$ produced in the collision also leads to recombination effects
\cite{BraunMunzinger:2000px,Thews:2000rj,Gorenstein:2000ck,Grandchamp:2001pf}
 which also obscure the use of these 
states as medium probes. Therefore, while $J/\psi$ states suffer a strong suppression in hadronic collisions \cite{Gonin:1996wn,Abreu:1997jh,Adare:2006ns,Aad:2010aa,Abelev:2012rv,Chatrchyan:2012np}, the suppression pattern is still puzzling; a  particularly surprising example of these puzzles is the  recent preliminary data  of $\psi (2S)$ production in Pb-Pb collisions at the LHC \cite{HPtalk}
which show an abnormal enhancement of $\psi (2S)$ relative to $J/\psi$ states,  which seem at odds with the sequential dissociation of the different quarkonia levels. 
Fortunately, data on bottomonium suppression have recently become available 
both at the LHC \cite{Chatrchyan:2012np} and at RHIC \cite{Reed:2011fr} opening a much more theoretically controlled channel. 

The theoretical description of in-medium quarkonia  properties has also experienced significant progress in recent years. 
Contrary to early expectations, current lattice analyses of charm and bottom current-current correlators indicate that quarkonia
states survive the deconfinement phase transition up to a temperature slightly higher than $T_c$ 
\cite{Asakawa:2003re,Datta:2003ww,Morrin:2005zq,Jakovac:2006sf,Aarts:2007pk,Aarts:2011sm,Ding:2012sp}.  
Although the determination 
of chamonium and bottomonium spectral functions is  still under intense theoretical investigation, there is a consensus that, at least,
the bottomonium ground states survive to a temperature as high as $T= 2\, T_c$, which is comparable to the highest temperatures reached at the LHC, while other excited states dissolve at a lower temperature. The situation for the in-medium $J/\psi$ is less clear, but indications from the lattice suggest that it is totally dissolved at $T<1.5 \, T_c$ \cite{Ding:2012sp}.

Complementary to lattice studies, potential models have been extensively used to describe the properties of finite-temperature quarkonia 
states
\cite{Karsch:1987pv,Digal:2001ue,Shuryak:2004tx,Wong:2004zr,Alberico:2005xw,Mannarelli:2005pz,Mocsy:2005qw,Mocsy:2007yj,Mocsy:2007jz,Cabrera:2006wh,Laine:2007gj}. 
In this approach the interactions among the heavy quarks in the bound states are encoded by an inter-quark potential and the different estates correspond to the non-relativistic levels of the corresponding  Schr\"odinger equation. 
In the vacuum, since the potential can be directly extracted from lattice calculations \cite{Kaczmarek:2005ui}, 
this approach provides a good description of the different vacuum quarkonia states \cite{Eichten:1979ms}. 
 In the medium, the main assumption is that all the interaction of the heavy pair with the plasma degrees of freedom can be codified in a medium-modified potential. However, contrary to the vacuum case, this temperature dependent  potential is not easy to extract from the lattice \cite{Laine:2006ns,Beraudo:2007ky} and most studies rely on phenomenological potentials adjusted to describe lattice correlators. In recent years, effective field theory approaches have put this potential description in firm theoretical grounds \cite{Escobedo:2008sy,Brambilla:2008cx,Brambilla:2010vq}, clarifying the region of 
validity of the potential assumption  at finite temperature. The explicit determination of the potential at sufficiently high temperature, such that the plasma dynamics are perturbative, have, in fact, led to the unexpected conclusion that in-medium potentials develop imaginary parts as a result of the destruction by the plasma of singlet $Q-\bar Q$ contributions \cite{Laine:2006ns,Beraudo:2007ky}.  

The in-medium properties of quarkonia extracted from the static studies above serve as an input for dynamical models of quarkonia production in 
ultra relativistic collisions. Over the last years  the description of $J/\psi$ production in these collisions has led to a  great body of work which accounts for a large variety of physical processes of relevance to this production (see \cite{Rapp:2008tf} and references therein for a comprehensive review). 
 More recently, the LHC capabilities for a detailed study of the  $\Upsilon$ family suppression have led to several studies focussed on these mesons \cite{Sharma:2009hn,Strickland:2011mw,Emerick:2011xu,Song:2011nu}.
In this work I will also address the suppression of heavy mesons by focussing on the effect a temperature dependent potential has on 
the production of quarkonia states in a deconfined hadronic matter. 
For simplicity, I will restrict myself to a deconfined medium with a fixed temperature which interacts with a heavy $Q-\bar Q$ pair during a
finite time (QGP-brick).  The pair is generated by a generic hard process and I will assume that the production is dominated by color singlet configurations. Contrary to other approaches, I  will not rely on adiabatic approximations for the in-medium states, but I will address the suppression dynamically by solving the time dependent Schr\"odinger equation without assuming the dominance of any in-medium state. 

The paper is organized as follows:  In section  \ref{potential} I describe the two simplified sample potentials which I used in substitution of the unknown  in-medium potential in the region of interest and which mimic some of the main features expected from this potential. In section \ref{brickmodel}
 I describe the QGP-brick set-up. The  formalism I have used to connect the hard production to the solutions of the time-dependent Schr\"odinger equation is described in 
subsection \ref{production} and   in subsection \ref{brick}  I  solve numerically  the 
time dependent Schr\"odiger equation and compare the quarkonia production in vacuum and in medium for 1S and 2S states.  Subsequently, in sections \ref{expansion}
and  \ref{RAA}, having in mind heavy ion physics application, I study the effect of a time and spatial variation of the in-medium potential. Finally, I discuss the main findings of this study in section \ref{conclusions}.

\section{Two Simple Models for the in-Medium Potential}
\label{potential}

Potential models have long been used to describe the properties of different quarkonia states in QCD plasma
\cite{Karsch:1987pv,Digal:2001ue,Shuryak:2004tx,Wong:2004zr,Alberico:2005xw,Mannarelli:2005pz,Mocsy:2005qw,Mocsy:2007yj,Mocsy:2007jz,Cabrera:2006wh}.
As already mentioned, the main assumption of this approach is that the interactions of the $Q-\bar Q$ pair with the thermal medium can be recast into a temperature dependent in-medium potential. While in recent years effective theory methods have led to a theoretical justification of this phenomenological approach in a certain limit ($1/r_B \sim \mu_D \gg E_b$, with $r_B$ and $E_b$ the Bohr radius and binding energy of the state and $\mu_D$ the Debye screening length of the plasma) \cite{Escobedo:2008sy,Brambilla:2008cx,Brambilla:2010vq}, the functional form of the thermal potential for temperatures slightly higher than $T_c$ remains unknown. Several studies based on phenomenologically motivated potentials have led to a remarkably good agreement with lattice data on heavy quark correlation functions
\cite{Mocsy:2007yj,Mocsy:2007jz,Cabrera:2006wh}. In most of these analyses, the in-medium potential is taken to be real and somewhat related to the singlet free energy computed via lattice QCD. However, in recent years perturbation theory analyses have shown that the in-medium potential does not need to be real since the interaction with the colored medium can change the color state of the pair, effectively reducing the probability of the pair to remain in a color singlet state \cite{Laine:2006ns,Beraudo:2007ky}.
While for the region of interest the perturbative expression for the potential is not reliable since in the vicinity of $T_c$ the coupling is not small, the many body effect described above exist beyond a perturbative treatment  and leads to  an imaginary part for the singlet potential, which leads the disappearance of singlet states in plasma. 

Given the absence of  an explicit form of the in-medium potential for the region of interest, in this work I will focus on the effect on quarkonia production of 
two toy-model potentials which mimic some of the most important features of the in-medium dynamics without any attempt to reconcile these potentials to lattice computations of the spectral function. Since this approach will already prevent me from a direct comparison with quarkonia suppression data on heavy ion collisions, I will further assume, for simplicity, that the vacuum quarkonia states are described by coulombic wave functions; while this is a reasonable approximation for sufficiently heavy quarks, significant deviations are found for realistic quark masses. 

\begin{figure}[tbp]
\centering 
\includegraphics[width=.45\textwidth]{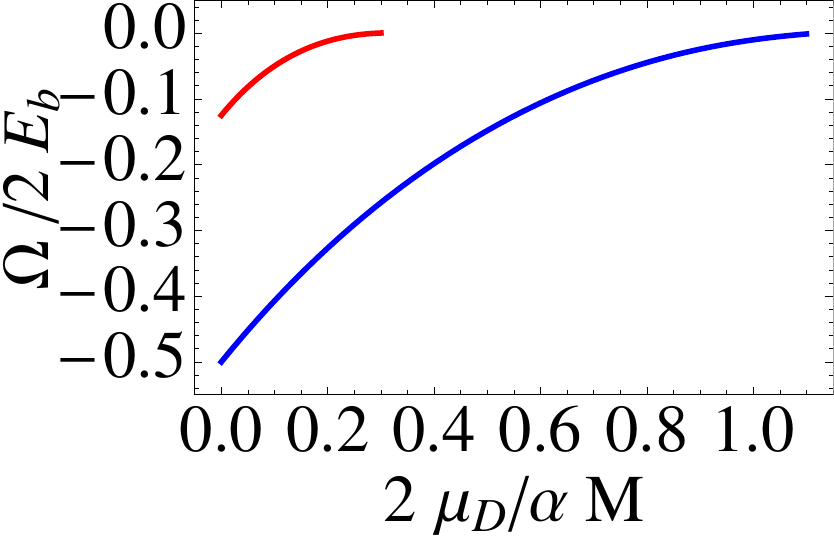}
\hfill
\includegraphics[width=.45\textwidth]{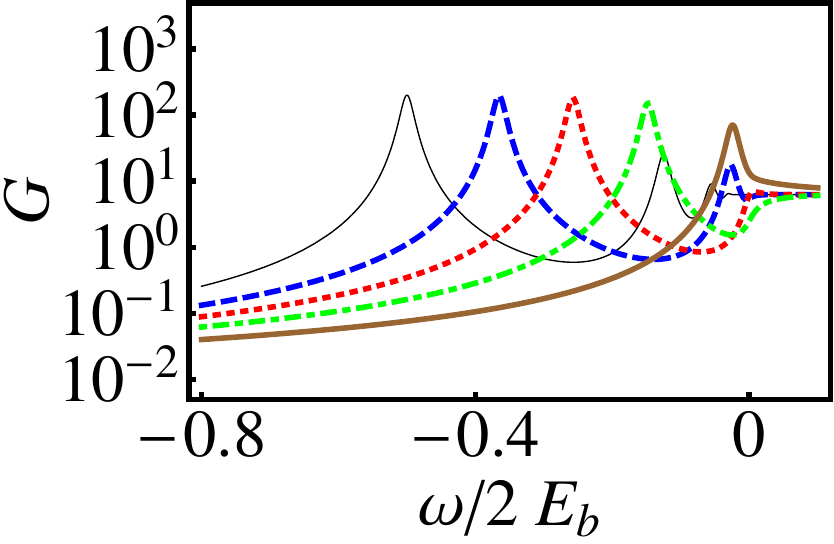}
\caption{\label{spRe} Left: Binding energy of the 1S (lower curve) and 2S (upper curve) states of the Yukawa potential, eq. (\ref{yukpot}), in units of the vacuum ground state anergy, $E_b=\alpha^2 M/4$ as a function of the Debye screening length $\mu_D$.
Right: Spectral function associated to the Yukawa potential  (in arbitrary units) as a function of the energy $\omega$, for different values of  $2 \mu_D/\alpha M=0,\, 0.15,\, 0.3,\, 0.5, \, 0.9 \,$ for the thin, dashed, dotted, dash dotted and solid lines respectively.}
\end{figure}

Since the relative magnitude of the imaginary part of the potential in the vicinity of $T_c$ is currently unknown, I will explore a (real) Yukawa potential, which encodes the color screening effects in the plasma. 
\be
\label{yukpot}
V_R(r)=-\alpha \frac{e^{-\mu_D r}}{r} \, ,
\ee
with $\mu_D$ the Debye screening length of the plasma, which characterizes the thermal state and $\alpha$ an adjustable parameter.  This potential coincides with the real part of 
the perturbative potential, up to a constant which is irrelevant for the temporal evolution. According to our assumption on the vacuum wave functions, we will consider the coulomb potential $\mu_D=0$ as the vacuum potential.

The  screening  of the quark color charge in the plasma  leads to the disappearance of the in-medium quarkonia states.
By solving the eigenvalue problem for different values of $\mu_D$, the biding energy of the 
ground state and the first excited state are determined. Their binding energies  as a function of $\mu_D$ are shown left panel of fig. (\ref{spRe}). According to  general expectations, excited states dissolved
at lower plasma temperatures than the ground state. 
 In the right panel of fig. (\ref{spRe}) I show the spectral function $G$ of the corresponding Schr\"odinger equation (see \cite{Mocsy:2007yj} for details on how to compute this spectral function). For illustration porpoises, we have evaluated it at $\omega=E-i \epsilon$ with 
$\epsilon =0.02 E_b$ and $E_b=\alpha^2 M /4 $ the binding energy of the vacuum ground state. At zero temperature (thin solid line), the spectral function shows several energy levels (the higher n excitations overlap because of the finite value of $\epsilon$). As the Debye screening length increases, the excited states disappear and the ground state moves towards threshold.  The reduction of the biding energy is accompanied by a reduction of the spectral strength, which coincides with a reduction of the modulus of the in-medium wave function at the origin.  Quarkonia states are, thus, expected to be suppressed in the medium even if they remain bound at finite temperature.

In order to study the effect of a complex in-medium potential, I also analyze a potential which differs from the Yukawa one, eq.~(\ref{yukpot}), by an imaginary part  
\be
\label{Vim}
V_I(r)=-\alpha \left(\frac{e^{-\mu_D r}}{r} +i \mu_D \,\phi (\mu_D r)\right) \, ,
\ee
with 
\be
\phi (r)=\int_0^\infty dz \frac{z}{(z^2+1)^2}\left(1-\frac{\sin( z r)}{z r}\right) \, ,
\ee
which coincides with the functional form of the imaginary part of the perturbative potential  \cite{Laine:2006ns,Beraudo:2007ky}. 
In comparing with the potential in  \cite{Laine:2006ns,Beraudo:2007ky} the reader may notice a different normalization of the imaginary contribution in eq.~(\ref{Vim}).
While in perturbation theory the prefactor of  $\phi$ is given by the temperate scale, T, in this work, for simplicity of the analysis, I have  used
 $\mu_D$ instead. 
Since in the region of interest we expect $\mu_D \lesssim T$, this expression will give a reasonable estimate of the effect of the imaginary part
on quarkonia suppression. In the rest of this paper, I will always assume that $\mu_D$ is of order $T$ and use one or another simultaneously.

\begin{figure}[tbp]
\includegraphics[width=.45\textwidth]{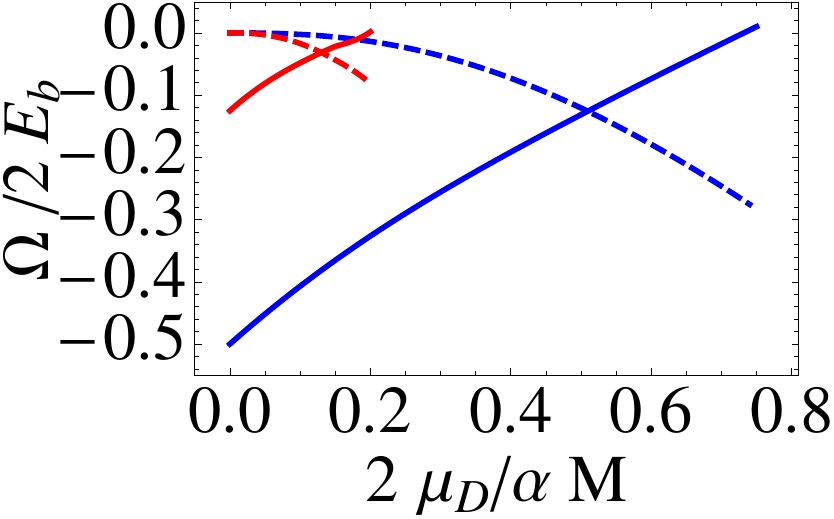}
\hfill
\includegraphics[width=.45\textwidth]{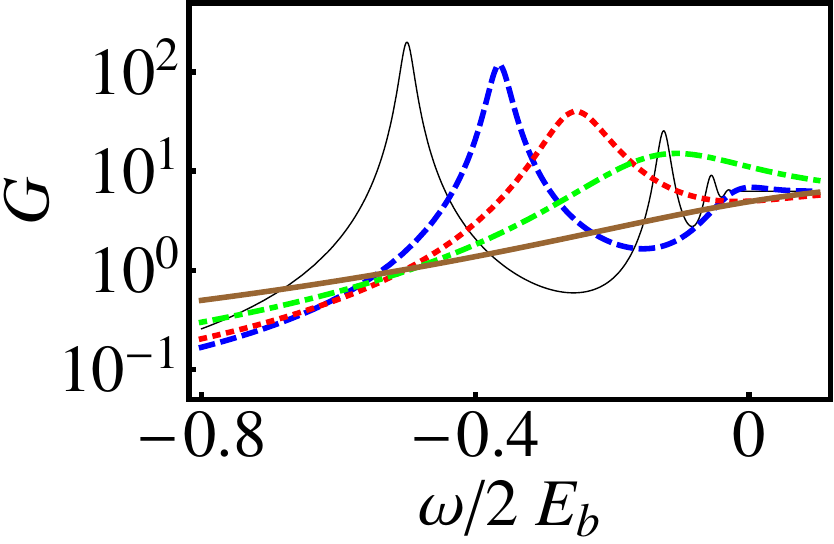}
\caption{Left: Real (solid) and imaginary (dashed) part of the binding energy of the 1S (lower curve) and 2S (upper curve) states of the complex potential, eq. (\ref{Vim})m in units of the vacuum ground state anergy, $E_b=\alpha^2 M/4$ as a function of the Debye screening length $\mu_D$.
Right: Spectral function associated to the same potential  (in arbitrary units) as a function of the energy $\omega$, for different values of 
$2 \mu_D/\alpha M=0,\, 0.15,\, 0.3,\, 0.5, \, 0.9 \,$ for the thin, dashed, dotted, dash dotted and solid lines respectively. }
\label{spIm}
\end{figure}

Thermal effects for this potential are twofold. The screening of the quark color charge leads to the disappearance of bound levels as in the previous
case; additionally, the complex potential introduces an imaginary part to the poles of the spectral density.
Thus, the eigenvalue problem for the potential in this case leads to complex eigenstates; in the right hand side of fig. (\ref{spIm}) I show the real and imaginary part
of the ground and first excited states. The  comparison of  the real part of the biding energy with those of the real potential,  fig. (\ref{spRe}), shows 
that the complex potential is more effective in  dissolving the bound states; additionally, the imaginary part of the eigenstates grows and becomes as large as the real one well before the bound state is dissolved. These imaginary eigenvalues are reflected in the broadening of the peak
structures in the spectral density 
shown in fig. (\ref{spIm}).  The spectral density also shows that, in addition to the faster level disappearance, the spectral strength of the state is reduced faster for the imaginary potential than for the real one; thus, we expect that the complex potential eq.~(\ref{Vim}) will lead to a stronger 
quenching of quarkonia states than the real potential eq.~(\ref{yukpot}). 

\section{The QGP-Brick}
\label{brickmodel}

The spectral functions computed in the previous section encode  the distribution of quarkonia states at fixed temperature (or $\mu_D$) in 
equilibrium. If the medium described by the potentials eq.~(\ref{Vim}) and eq.~(\ref{yukpot}) would be sufficiently large and exist for a
sufficiently long time, the asymptotic distribution of quarkonia states would be a thermal distribution according to that spectral function and 
independent 
of the origin of the $Q-\bar Q$. Furthermore, for these long media, the dilepton rate in the region of invariant mass of order $2 M$ would be directly related to those spectral functions. 
However, for any realistic application the medium is finite and short-lived.  Thus, the study of not only the pre-asymptotic dynamics  but also the modification of the formation of the long-lived vacuum quarkonia  states by the presence of a finite
medium  are of relevance for the understanding of quarkonia production in a heavy ion environment. 

\begin{figure}[tbp]
\centering 
\includegraphics[width=0.8\textwidth]{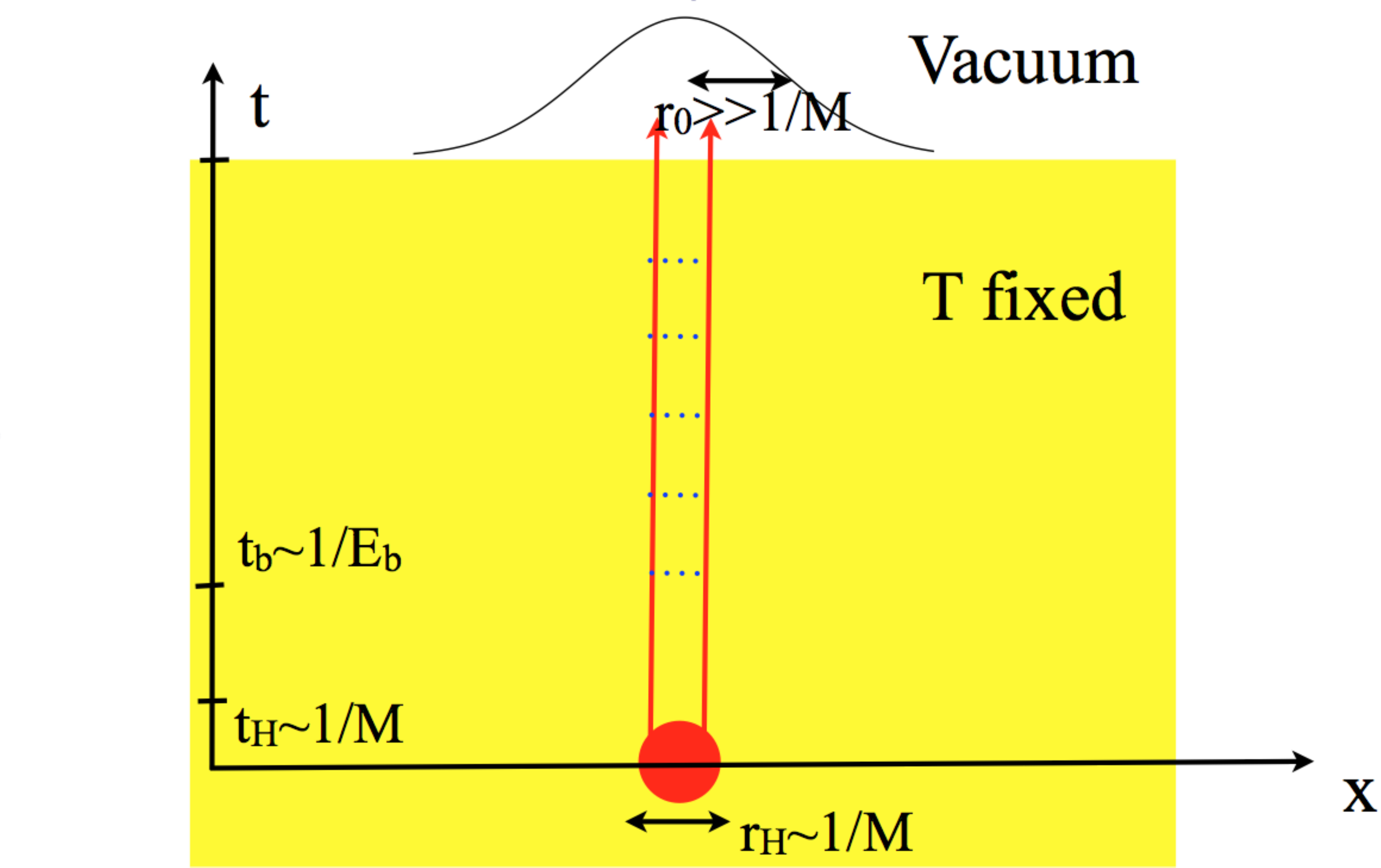}
\caption{Sketch of the QGP-Brick set up. At initial time a $Q-\bar Q$ pair is formed in the medium via a hard process that occurs within a short time $t_H\sim 1/M$. At a later time of order the inverse binding energy, the soft interactions lead to the formation of quarkonia states from the pair, which are influenced by the presence of the medium. At a later time t, the medium disappears and the $Q-\bar Q$ system projects into vacuum quarkonia states. } 
\label{sketch}
\end{figure}

To study those processes, I will first study quarkonia production in a homogeneous medium at fixed temperature with a finite lifetime after which the temperature drops abruptly to zero. The spacial extent of the medium is  taken to be large, much larger that the in-medium radius of bound states.  I will assume that a $Q-\bar Q$ pair is formed within the medium via a hard process at a time $t_0=0$ and interacts with it during a finite time $t$ after which the medium disappears. After this time, the evolution occurs as in vacuum.  
In analogy with the jet quenching  studies, where a similar set-up was used to analyze the differences among the available quenching models, I will call this set-up, which I have sketched in fig.~(\ref{sketch}), a QGP-brick.

\subsection{Quarkonia Production from a Hard Process}
\label{production}

I focus in the production of  quarkonia states from a hard process which takes  place within the QGP-brick. 
For sufficiently heavy quarks ($M_q\gg \Lambda_{QCD}$), this  process   can be factorized into a short distance behavior, described via perturbation theory, and a long distance, non perturbative matrix element. The production cross section of a state S is   expressed as
\cite{Bodwin:1994jh} 
\be
d \sigma (S)=\sum_{n} d\sigma (Q \bar Q ([n])) \left<Q \bar Q ([n]) \rightarrow S\right> \, ,
\ee
where $d\sigma (Q \bar Q ([n]))$ is the perturbative production cross section of a  $Q-\bar Q$ pair with quantum numbers $n$ and  $\left<Q \bar Q ([n])  \rightarrow S\right> $ is the soft matrix element which projects a given  $Q-\bar Q$ pair into the state S.  The set of quantum numbers $n$
does not need to coincide  with the quantum numbers of the final state.  However, for sufficiently heavy quarks, for  which the relative velocity v of the $Q-\bar Q$ pair is very small, the contribution of those configurations with quantum numbers different from those of the state S are suppressed by powers of  $v$.  Therefore, in the rest of the paper I will neglect those corrections \footnote{These include the contribution of color octet states; I am, thus, restricting myself to a color singlet model of quarkonia production.}. In the particular case of vector S-wave states (such us $J/\psi$ or $\Upsilon$) the production cross section is reduced to 
\be
d\sigma (S)\approx d\sigma (Q \bar Q ([{}^3S_1])) \left<Q \bar Q ([{}^3S_1]) \rightarrow S\right>  \, .
\ee

I will also assume that the heavy quarks are much heavier than any medium scale, in particular, $M\gg T$.  Thus, 
since the hard part of the production cross section  takes place within a space-time region of typical size $1/M$, which is, by assumption, much smaller than any medium scale, the presence of the plasma cannot alter this part of the production process, and $ d\sigma (Q \bar Q ([{}^3S_1])) $ remains the same irrespectively of wether  the production takes place in the vacuum or within a medium.\footnote{With this approximation I am also neglecting quark energy loss effects on  quarkonia production. While these do not change significantly the production rate, they can alter the momentum distribution of the observed states.} 
Within this approximation, plasma effects are only present in the late time  projection of the $Q-\bar Q$ pair   into the particular quarkonia state. For sufficiently heavy quarks,  potential non-relativistic QCD (pNRQCD), which describes the dynamics of soft $M v^2$ modes can be used to determine these  matrix elements. 
In this approach, the quarkonia  dynamics are described via the solution of the  Schr\"odinger equation with a given inter-quark potential and the matrix element is determined from the $Q-\bar Q$ pair wave function.  

For central $Q-\bar Q$ potentials, the center of mass dynamics of the pair decouples and the matrix element is only dependent on the equivalent one body problem. 
From the point of view of the soft dynamics, the hard production can be  described via a Wigner function for the $Q-\bar Q$ pair, $W_H\left({\bf r}, {\bf q}\right)$, with 
${\bf r}$ and $ {\bf q}$ the relative  position and momentum of the pair at an initial time $t_0$ within the box.  Due to the uncertainty principle, $t_0$
cannot be determined with an accuracy larger than $1/M$ but, since I have assumed that this scale is much smaller than any medium or bound state scale, I will neglect this uncertainty. From the Wigner function, the initial one body density matrix is given by 
\be
\rho_H\left({\bf r}-\frac{\bf y}{2},{\bf r}+\frac{\bf y}{2}\, ; t_0\right)=\int \frac{d{\bf q}}{(2\pi)^3} e^{i {\bf q\,{\bf y}}} W_H\left({\bf r}, {\bf q}\right) \,. 
\ee

Independently of the particular process that leads to the production of a $Q-\bar Q$  pair within the medium, since the process is hard, the typical momentum of each of the quarks, $p$, is large $M\lesssim p$. Thus, the typical  relative quark momentum is also   
much larger than the inverse bound state radius  and, from the point of view of the soft matrix element, 
only the $q\rightarrow 0$ limit of 
 $W_H$ is relevant for the production of bound states, which leads to an approximate  initial density matrix given by 
\be
\label{rhohard}
\rho_H\left({\bf r}-\frac{\bf y}{2},{\bf r}+\frac{\bf y}{2}\, ; t_0\right)=\rho_0({\bf r}) \delta \left({\bf y}\right)\, .
\ee
Furthermore, 
since the hard process takes place at very small distances I will approximate the function $\rho_0({\bf r})$ by a   $\delta$-function at the origin,  $\rho_0({\bf r})\propto \delta ({\bf r})$.

As the $Q-\bar Q$ pair propagates,  the density  matrix eq~(\ref{rhohard})  evolves via a non-relativistic hamiltonian, since the late time interactions are soft as compared to the pair mass. At any given time $t>t_0$ the different particle yields can be  obtained by projecting the  evolved density matrix into the single state density matrix
\be
\label{yieldgen}
Y_S \propto {\rm Tr} \left(\rho_H(t) \rho_s \right)={\rm Tr} \left(\rho_H(t_0) \rho_s(t-t_0) \right)\,,
\ee
with $ \rho_s=\left| S \right> \left< S\right|$ and $\left| S \right>$ the quarkonium estate of interest. In the second equality above I have used the cyclicity of the trace to define 
\be
\label{rhoev}
 \rho_s(t-t_0) = \left(e^{-i H (t-t_0)}\right)^\dagger\left|  S \right> \left< S\right| \left(e^{-i H (t-t_0)}\right) =\left| \tilde S\right>\left< \tilde S\right| \, ,
\ee
where $\left|  \tilde S \right> $ is the (backwards in time) evolved wave function. Taking $t_0$ as the evolution variable and performing the 
change $\tilde t =t-t_0$ this wave function satisfies the  (hermitian conjugate) Schr\"odinger equation:
\be
\label{schrod}
- i\partial_{\tilde t} \tilde \psi (\tilde t,{\rm r})= -\frac{\nabla^2}{M_q} \tilde \psi(\tilde t,{\rm r})+ V^\dagger(r) \tilde \psi (\tilde t,{\rm r}) \,,
\ee
with $\tilde \psi (\tilde t,{\rm r})=\left< {\bf r} | \tilde S \right>$ and initial condition $\tilde \psi (0,{\rm r})=\left< {\bf r} | S \right>$.

Combining eq.~(\ref{rhohard}), eq.~(\ref{yieldgen}) and eq.~(\ref{rhoev}), 
the yield of quarkonia production after a time of propagation t is given by 
\be
Y_S \propto \left| \tilde \psi(t -t_0,{\bf r}=0) \right|^2 \, .
\ee

In the vacuum, the inter quark potential is hermitian and the modulus of the the wave function at the origin remains constant as time changes. Thus the expression above coincides with the standard matrix elements for the computation of quarkonia states \cite{Bodwin:1994jh}.  For the production in the medium-brick, the situation is more complicated, since the interaction with the QGP fields changes the potential during the time extent of the medium; furthermore, when the $Q-\bar Q$ leaves the QGP-brick, the potential changes abruptly back to the vacuum expression. If all the thermal effects on the  $Q-\bar Q$  system can be expressed in terms of a medium-modified potential, the complicated dynamics of the production process in this set up can be addressed by solving the time dependent Schr\"odinger equation, eq.~(\ref{schrod}) once the explicit
form of the potential is known.

\subsection{Quarkonia Suppression in a QGP-Brick}
\label{brick}

Since after the particle leaves the brick the yield of the different states remains constant, in order to determine the production of quarkonia in the system it is enough to compute the evolution of the hard density matrix up to the time $t$ when the medium disappears.  Following the previous discussion, the yield of a given state  is obtained by   (numerically) solving the time dependent Schr\"odiger equation, eq~(\ref{schrod}), with the in-medium potential, eq.~(\ref{Vim}) and eq.~(\ref{yukpot}),  starting at $\tilde t=0$ with its vacuum wave functions and computing the value of the evolved wave function at the origin at time $t$. From this solution I determine the 
 yield suppression by the medium, which I express,  as customary in heavy ion physics, as the ratio of the yields after the QGP-brick to the one in vacuum 
\be
\label{Rdef}
R_S=\frac{Y_S(\mu_D)}{Y_S(\mu_D=0)} \,.
\ee

\begin{figure}[tbp]
\includegraphics[width=.45\textwidth]{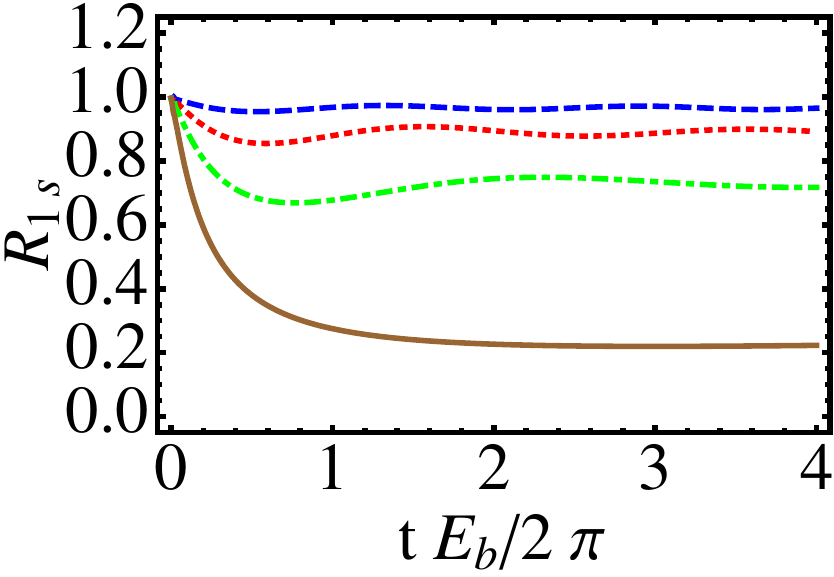}
\hfill
\includegraphics[width=.45\textwidth.]{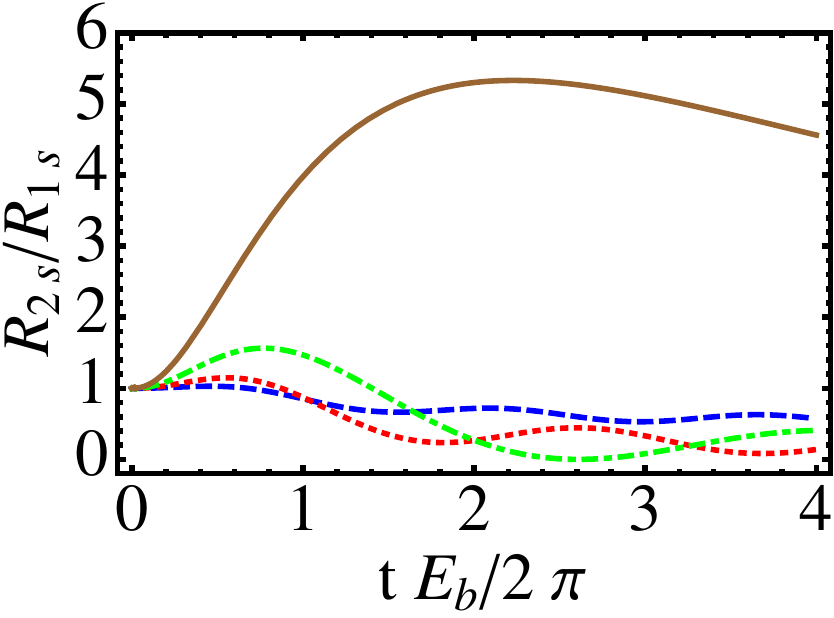}
\caption{Left:  Yield suppression of the ground state level as a function of the lifetime of the QGP-brick, t, in units of the vacuum ground state period $\mathcal{T}=2 \pi/ E_b$. Right: Double ratio of the suppression factors of 2S and 1S states as function of the lifetime of the QGP-brick. Both calculations are for the real potential eq~(\ref{yukpot}) with $2 \mu_D/\alpha M=0.15,\, 0.3,\, 0.5, \, 0.9 \,$ for the dashed, dotted, dash dotted and solid lines respectively. } 
\label{SupRe}
\end{figure}

In the left panel of fig.~(\ref{SupRe}) I show the suppression of the  S-wave ground state (1S) for different values of $\mu_D$  as a function of the time duration of the box in units of the inverse period of the ground state wave function $\mathcal{T}=2\pi/ E_b$, which I will identify as the  formation time.  For very small brick lifetimes, less than one period, the suppression is relatively modest for all values of $\mu_D$. At later times and as long as the in medium potential supports a bounded ground state, the yield suppression saturates to a constant value, which is determined by the value of the in-medium wave function at the origin and the overlap of those wave functions with the vacuum ground state. 
As $\mu_D$ grows, the late time suppression also grows, since the in-medium states becomes wider as they approach threshold, reducing both the value of their wave function at zero and the overlap with the vacuum wave function. When $\mu_D$ is so large that the ground states disappears 
(not shown) the late time yield tends to zero.
 
 I have also performed a similar analysis for the suppression of the next vacuum state, the 2S level. In the right panel of Fig~(\ref{SupRe}) 
 I show the double ratio of the suppression factor of the 2S state to the 1S estate. Note that for only one of the cases considered, $\mu_D=0.15 \alpha M/2$ (dashed line in fig.~(\ref{SupRe}) )
 the in-medium potential supports 2S bound level and the medium modifications are small. As the temperature rises,  a modulation in the 
 double yield ratio appears. The physical origin of this is simple and it is due to the non-zero overlap of the vacuum 2S state with the in-medium ground state, which for this potential survives for all the shown values of $\mu_D$.  The period of oscillation is given by the in-medium ground state energy and the amplitude is controlled by the interference between the in-medium continuous states and the ground state, which disappears at later times since the former states are delocalized.  Quite remarkably, this double ratio can be larger than 1, showing a relative enhancement of 2S to 1S 
 states. This is particularly dramatic in the highest temperature case, for which the bound states are loosely bound.  This behavior continues at even higher temperatures, when all states are dissolved and it is due to the fact that the vacuum 2S state is wider than the ground state and has a larger overlap with the in-medium continuum. However, I emphasize that despite the relative enhancement, in this situation both the 1S and 2S states are   strongly suppressed  as compared to the vacuum.

\begin{figure}[tbp]
\includegraphics[width=.45\textwidth]{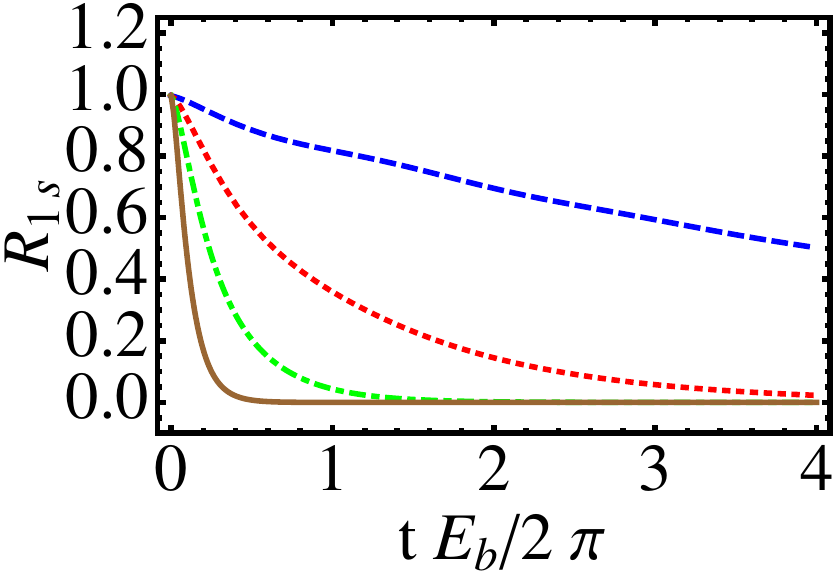}
\hfill
\includegraphics[width=.45\textwidth]{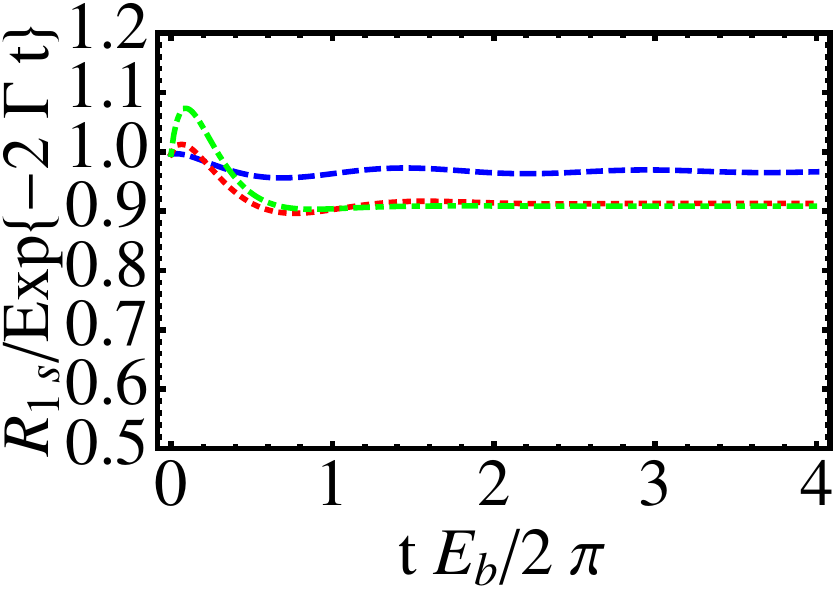}
\caption{Left:  Yield suppression of the ground state level as a function of the lifetime of the QGP-brick, t, in units of the vacuum ground state period $\mathcal{T}=2 \pi/ E_b$ for the complex potential eq~(\ref{Vim})  for different values of $\mu_D$,  $2 \mu_D/\alpha M=0.15,\, 0.3,\, 0.5, \, 0.9. $  Right: Comparison of the suppression factor to the attenuation of the yield expected from the imaginary part of the lowest in-medium mode for 
the previous values of $\mu_D$ which support an in-medium state.} 
\label{SupIm}
\end{figure}

The introduction of an imaginary part to the potential leads to some distinct features in the suppression pattern  of the QGP brick.
As shown in the left panel of fig.~(\ref{SupIm}), for the same values of $\mu_D$, the complex potential, eq.~(\ref{Vim}) is much more effective in 
suppressing quarkonia. Contrary to fig.~(\ref{SupRe}), for this potential  the late time suppression does not saturate to a constant, but continuously 
increases due to the finite width of the in-medium states. In fact, at these late times, the dynamics of the in-medium ground state is dominated by this absorption.  This is demonstrated in the right panel of fig.~(\ref{SupIm}), where I show the ratio of the numerically computed suppression divided by the expected absorption rate due to the width $\Gamma$ of the in-medium ground state for the values of $\mu_D$ for which it survives.  As shown by 
the plot, after the formation time of the meson the absorption is indeed given by the in-medium width up to an overall shift, which is due to the early time suppression.

\begin{figure}[tbp]
\includegraphics[width=.45\textwidth]{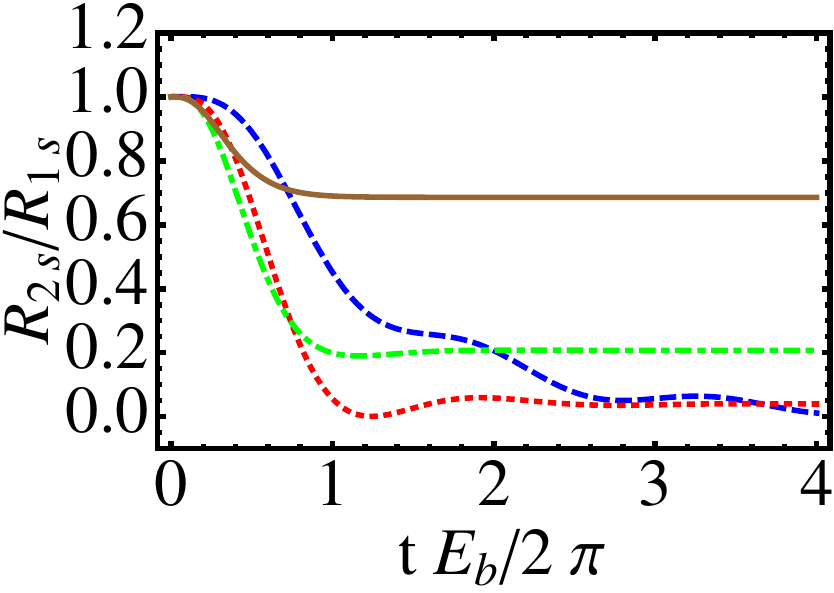}
\hfill
\includegraphics[width=.45\textwidth]{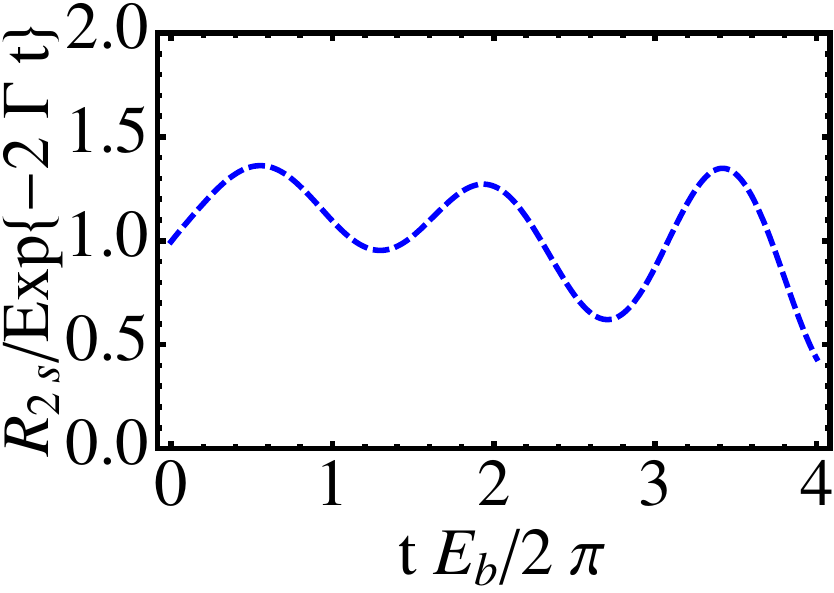}
\caption{Left:  Double ratio of the suppression factors of the 2S and 1S states as a function of the lifetime of the QGP-brick, t, in units of the vacuum ground state period $\mathcal{T}=2 \pi/ E_b$ for the complex potential eq~(\ref{Vim})  for different values of $\mu_D$,  $2 \mu_D/\alpha M=0.15,\, 0.3,\, 0.5, \, 0.9. $  Right: Comparison of the suppression factor to the attenuation of the yield expected from the imaginary part of the 2S in-medium mode for 
the only value of $\mu_D$ which supports a bound 2S state.} 
\label{R2s21sIm}
\end{figure}

In the left panel of fig.~(\ref{R2s21sIm}) I show the double ratio of suppression of the 2S to 1S state for this imaginary
potential.  
For the three highest values of 
$\mu_D$  shown, for which the potential does not support a 1S level, the late 
 time ratio is constant, which means that both the 2S and 1S yields fall at the same rate.  Similarly
 to the real potential case, this is due to the fact that when the in-medium 2S level is absent, 
 the production is dominated by the overlap with the in-medium 1S.  Once again, the typical 
 time scale to reach to this asymptotic behavior is given by the ground state period. As in the real potential
 case, the asymptotic value grows with $\mu_D$. Even though the double ratio does not go above one
 for any of the shown values, I have checked that further increasing $\mu_D$ leads to an enhancement over
 the vacuum, as in the previous case, but such that each individual yield is strongly suppressed.
 
  The effect of the observed
 level overlap is also present for colder media, in which a 2S level exist. In the right panel of 
 fig.~(\ref{R2s21sIm})  I compare the numerically computed suppression factor to the absorption rate due
 to the imaginary part on the in-medium level. Contrary to the ground state, the rate is not saturated by the 
 in-medium level, and there are large oscillations around this decay. The period of those agrees, in fact, with the 
 real part of the in-medium 1S level. Furthermore, since the imaginary part of the 
 1S level is smaller than the 2S one, the ground state contribution grows at late times.

In summary, by comparing the suppression patterns of a medium-brick characterized by a real 
quarkonia potential to that of a potential with an imaginary part  with the same screening parameter $\mu_D$ I 
have shown  that the non-unitarity effects introduced by the complex potential are very efficient
in suppressing quarkonia. Even a relatively small imaginary part has dramatic consequences in the 
late time suppression of quarkonia, specially for long lived medium-bricks. On the contrary, for short 
lived bricks, with a lifetime less than the formation time, the differences between these two models of the medium are not
so dramatic. Additionally, I have shown that even though the ground state suppression pattern is easy to understand
in terms of in-medium bound states, the suppression of excited states, such as 2S-wave levels, have a much
richer structure which arises not only from the contribution of the corresponding 2S  in-medium level, but also
from the ground state, which in some cases become dominant.

\section{Quarkonia Suppression in an Expanding Medium}
\label{expansion}
The analysis of the static brick in the previous section, in which the medium parameters are fixed 
during the medium lifetime, shows that the in-medium suppression patterns develop in a characteristic
time of the order of  the  ground state period. A simple estimate of these times for chamonium and
bottomonium bound states yields  typical
formation times $\mathcal{T}$ of few fm, which
 having in mind any heavy ion physics applications, 
are large, comparable to the whole fireball lifetime.  Thus, in this section I will extend the analysis to an 
"expanding" brick, for which I mean a spatially infinite homogeneous medium whose properties change
continuously with time, till the medium completely disappears at a given time.  I will assume that the internal
processes in the medium are sufficiently fast to keep the medium thermalized during the expansion.
Motivated by the dynamics
of boost invariant plasmas, a common approximation in describing hydrodynamic evolution in heavy ion 
collisions, I will assume a time dependence of  $\mu_D$ given by
\be
\label{mudt}
\mu_D(t)=\mu_{D0}\left(\frac{\tau_0}{t + \tau_0}\right)^{1/3} \,,
\ee
with $\mu_0$ the initial value of the Debye screening length and $\tau_0$ a thermalization
time parameter which I will take to be $\tau_0=0.4$ fm. To mimic the deconfinement transition,
I demand that the brick lifetime coincides with the time in with $\mu_D=175$ MeV, 
the phase transition temperature, where I am assuming, as already mention in section \ref{potential}, that
in the vicinity of $T_c$, $\mu_D\approx T$. In a real expansion, the hadronic medium does not cease to exist
after the system reaches $T_c$. However, after this time the in-medium modifications of quarkonia are much smaller, since the medium is confined. In this simplified approach I am  neglecting the interaction of quarkonia states with the dense hadronic medium after the transition.

To  study the effect of the expanding and deconfined medium into  the different quarkonia states, I need to
set physical values for the wave function parameters. Since given the many assumptions of this exploratory  work I will not be 
able to make direct contact with experimental data,
 I will restrict myself to the coulombic wave functions used in the previous
analysis without questioning how approximate they are to those of a realistic potential. To fix
the scale, I will equate the r.m.s of the coulombic wave function, $r_{rms}=<r^2>^{1/2}$, to 
the one of wave function  computed with a realistic vacuum potential \cite{Eichten:1979ms,Mocsy:2005qw}. Using 
$r_{rms}= 0.5\,, 0.3$ fm for $J/\psi$ and $\Upsilon$ respectively,  the effective $\alpha$ is different for
charm and bottom  bound states,  $\alpha=1\,, 0.5$ respectively.  This value also sets the binding 
energies of both ground states, $E_{b}=0.32,\, 0.26$ GeV for $J/\psi$ and $\Upsilon$.
These binding energies are low as compared to the binding energies of vacuum $J/\psi$ and $\Upsilon$
states $E_b\approx 0.7,\,  1\, $ GeV, respectively; however, they are close to the in-medium binding 
energies right above $T_C$ extracted from potential model description of lattice heavy quark current-current
correlators \cite{Mocsy:2007jz,Cabrera:2006wh}, which is the relevant scale for describing the decorrelation among different in-medium
states. Nevertheless, because of this mismatch, all the time scale estimates in this work should be taken as
indicative.

\begin{figure}[tbp]
\includegraphics[width=.45\textwidth]{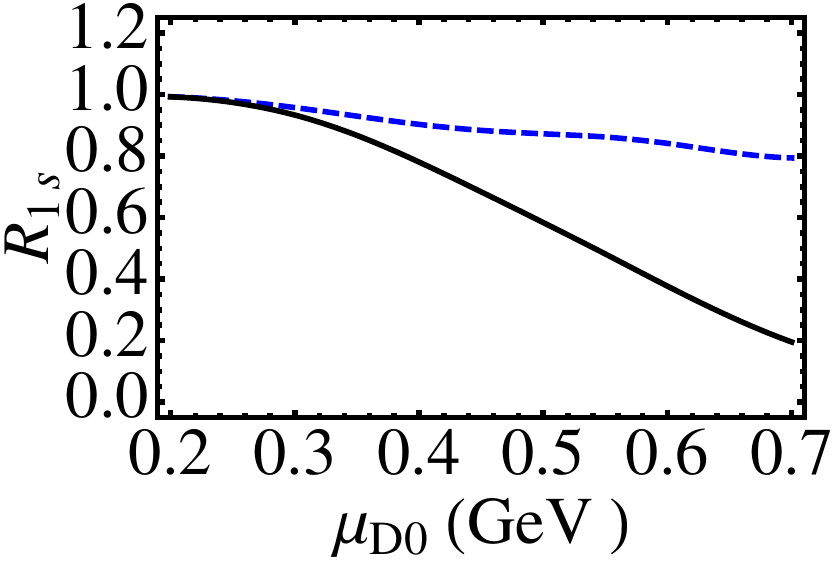}
\hfill
\includegraphics[width=.45\textwidth]{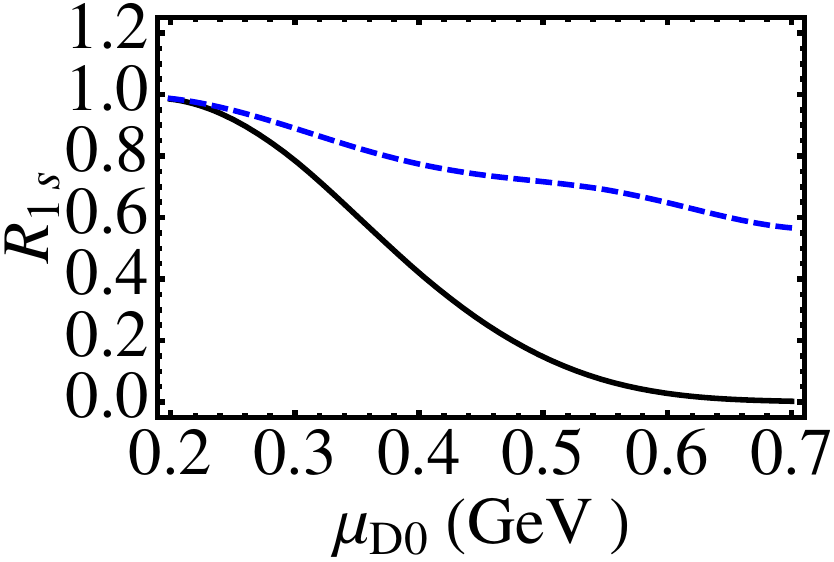}
\caption{Suppression factor of the 1S state in a finite lifetime expanding medium as a function of the initial value of the 
$\mu_D$ parameter for both the real  (dashed) and complex (solid) potentials. The left and right panels correspond to $\Upsilon$ and
$J/\psi$ suppressions respectively.} 
\label{R1sofmu}
\end{figure}

I now repeat the analysis in the previous section with the time dependent potential arising from 
the use of eq.~(\ref{mudt}) in eq.~(\ref{schrod})  and eq.~(\ref{Vim}).  As in the static case, I evolve the vacuum wave functions with 
 the Schr\"odinger   equation, eq.~(\ref{schrod}), up to the final time, defined by $\mu_D=T_c$. 
In fig.~(\ref{R1sofmu}) I show the suppression of the $\Upsilon$ (left) and $J/\psi$ (right) states 
both for real (dashed) and complex potentials (solid), as a function  of the initial screening length
$\mu_{D0}$. As expected from the analysis of the static medium, the complex potential is much
more effective in suppressing quarkonia states than the real one. While the real potential  
leads to a moderate suppression both for $J/\psi$ and $\Upsilon$, the complex potential leads to 
a larger  suppression with a stronger dependence on the initial $\mu_D$ value.                
Note that with the current choice of parameters, for the most part of the evolution the value of 
$\mu_D$ at any time is smaller than the dissociation value, inferred from fig.~(\ref{spIm}). The 
 suppression pattern observed is, in fact, a consequence of the 
transient behavior appearing at short times in the fixed $\mu_D$ calculations of the previous section, since  for $\mu_{D0}< 0.5$ GeV
the lifetime of the medium is smaller than $\mathcal{T}$.

\begin{figure}
\includegraphics[width=.45\textwidth]{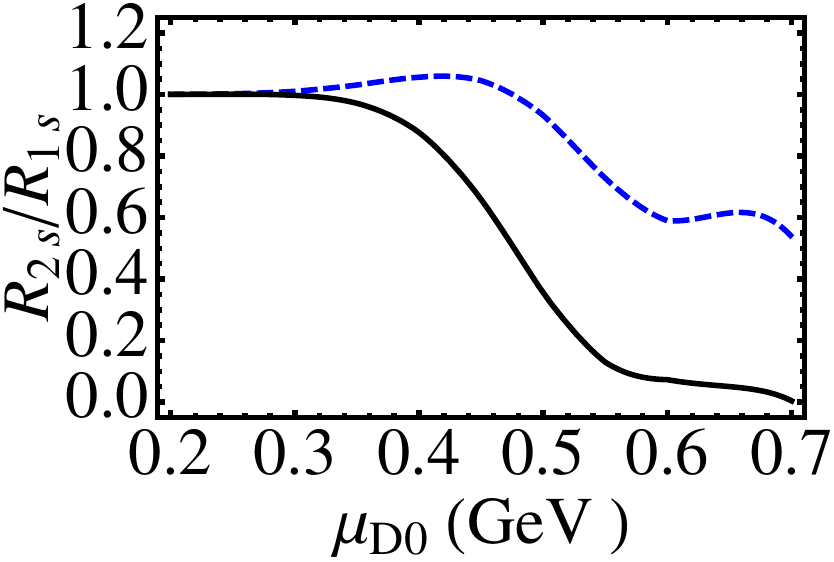}
\hfill
\includegraphics[width=.45\textwidth]{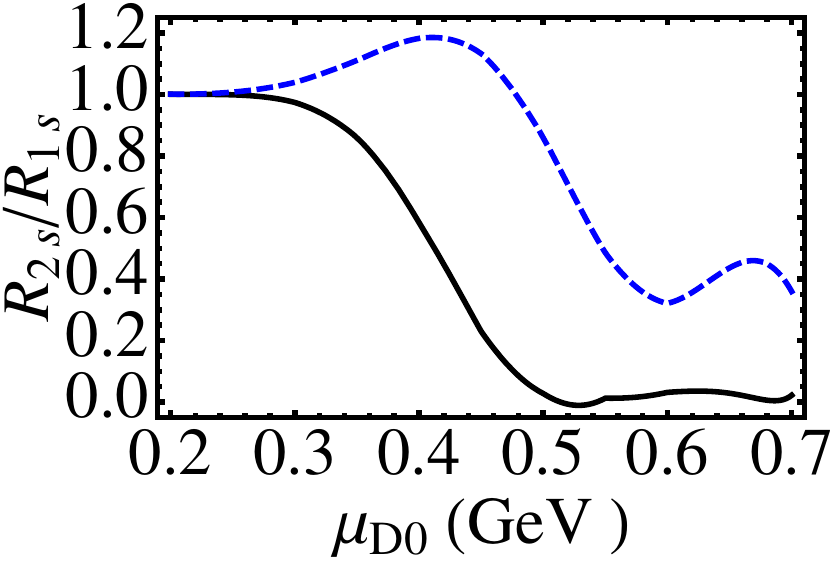}
\caption{Double ratio of the suppression factors of the 2S to 1S states in a finite lifetime expanding medium as a function of the initial value of the 
$\mu_D$ parameter for both the real  (dashed) and complex (solid) potentials. The left and right panels correspond to $\Upsilon$ and
$J/\psi$ suppressions respectively.} 
\label{R2s21sofmu}
\end{figure}

In fig.~(\ref{R2s21sofmu}) I show the double ratio of the suppression factors of $2S$ and $1S$ states.  Contrary to the fixed temperature
calculation, neither for the $\Upsilon$ nor $J/\psi$ case the double ratio is significantly larger than one, despite of the fact that, at least for $J/\psi$,
the highest values of $\mu_D$ are sufficiently large to expect an enhancement. The reason for this lack of enhancement is that even for the hottest initial medium, the fast evolution at early times makes $\mu_D$ to drop significantly within the transient time, $\mathcal{T}$. Thus, the final yield is dominated
by low values of $\mu_D$ for which, within the current parameter set,  the 1S states are bound. Note, however, that because of my simplified choice of in-medium potentials, quarkonia states survive till large values of the $\mu_D$, in contrast with other more realistic potential 
model calculations \cite{Mocsy:2007yj,Mocsy:2007jz}.
 Nevertheless, this calculation shows the insensitivity of the suppression pattern to the very early stages of the evolution, where the temperature is the highest, as a consequence of the finite formation time of the different quarkonia states. 

\section{Quarkonia Suppression in a non-Homogenous Expanding Medium}
\label{RAA}
Since there is a non-trivial relation between the static brick studied in section~(\ref{brick}) and the expanding case of the previous section, I
will now investigate the effect of a non-homogenous temperature profile in the final quarkonia production rate after a finite time in the 
hot medium. Having in mind ultra relativistic heavy ion collision applications, I will consider a medium of finite spacial extent with varying initial $\mu_{D0}$
which expands at a (local) rate given by eq.~(\ref{mudt}). In this study, the initial parameter $\mu_{D0}$ becomes position dependent; however
I will assume that the typical size of in-medium bound states is small as compared to the typical variation of the $\mu$-profile of the medium 
and neglect the spatial variation of the in-medium hamiltonian.

As a simplified model for the  medium formed after the collision of the two heavy nuclei I will use  a hard spheres model with a strong longitudinal expansion but no transverse dynamics. Right after the collision, there is a non-vanishing  temperature profile inside the overlap region of the two spheres.  At any impact parameter  $b$,  this  profile is given by 
\be
\label{mudT}
\mu_D=\mathcal{C}\left( T_A((x-b/2),y) +T_A((x+b/2),y)\right)
\ee
with $T_A(x,y)=\int \rho(x,y,z) dz$ the thickness function of the nuclei of density $\rho$ and $\mathcal{C}$ a proportionality constant. For this study, I will set
$\mathcal{C}$  by demanding that  in central collisions ($b=0$) the highest value of $\mu_D=0.7$ GeV.  This procedure assigns to every point in the transverse plane of the overlap region a non-vanishing value of $\mu_D$ which decreases as the point is further away from he center of that region. As in the previous sections I will assume that if  $\mu_D< 0.175$ GeV there is not a substantial modification of the in-medium potential and 
for all those points of the transverse plane for which eq.~(\ref{mudT}) yields values smaller than $T_c$ I set $\mu_D=0$. For all the other points, 
$\mu_D$ decreases with time as dictated by eq.~(\ref{mudT}).

Having determined the $\mu_D$ profile, I also need to specify the distribution of production points of the $Q\,- \bar Q$ pairs. Without specifying  the 
microscopic mechanism that leads to the creation of pairs in singlet state, I will assume that the probability of production scales with the 
number of binary collisions 
\be
\mathcal{P}\propto T_{AA} (x,y; b)
\ee
with $T_{AA} (x,y;b)=T_A(x-b/2,y) T_A(x+b/2,y)$ defined as usual.   I will also assume that the $Q\,-\bar Q$ pair is produced mostly at rest and leave the 
momentum dependence of the suppression for a future study. Thus, the effective hamiltonian governing the soft dynamics of the pair is controlled
by the local temperature at the production point and its subsequent time evolution.  Averaging over the overlap region I obtain the yield of the different states associated to this simplified model of nuclear collisions.

\begin{figure}[tbp]
\includegraphics[width=.45\textwidth]{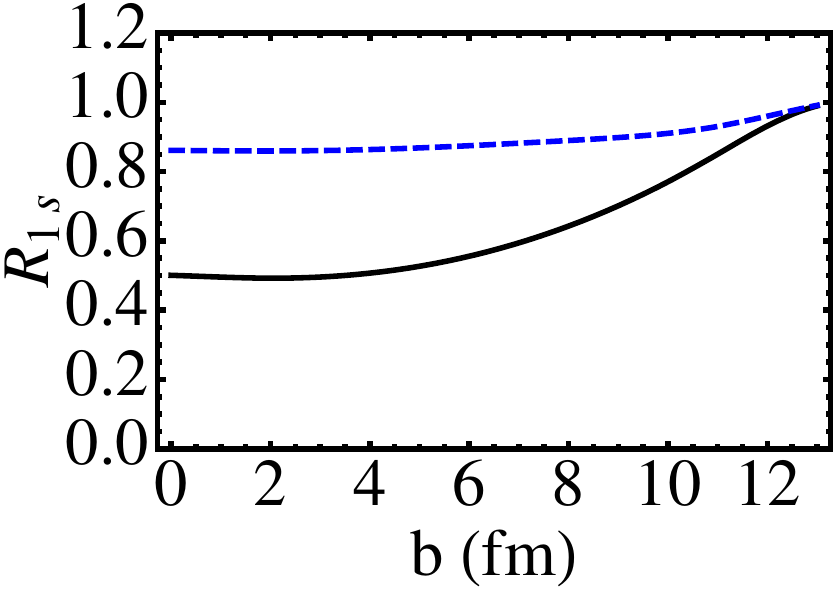}
\hfill
\includegraphics[width=.45\textwidth]{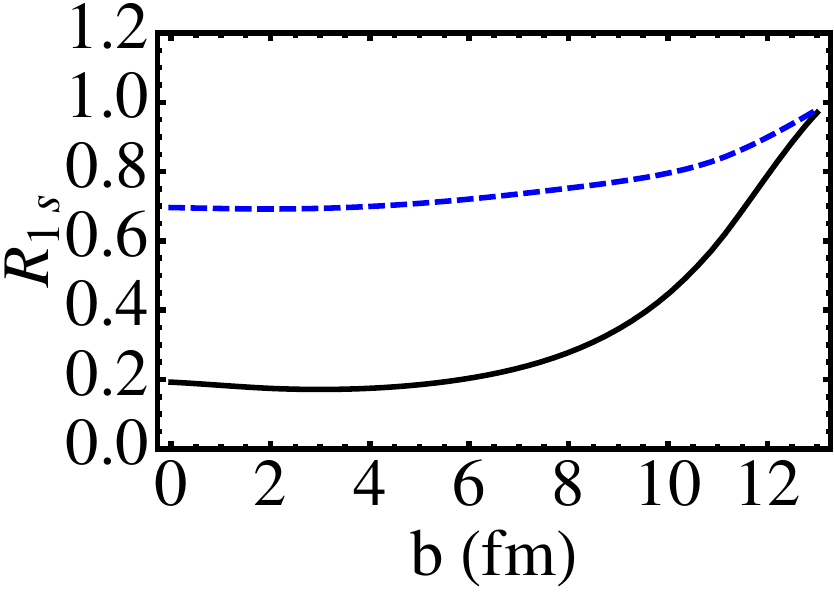}
\caption{Suppression factor of the 1S state for the spherical nuclei collisions as a function of impact parameter
$b$ for both the real  (dashed) and complex (solid) potentials. The left and right panels correspond to $\Upsilon$ and
$J/\psi$ suppressions respectively.}
\label{R1sofb}
\end{figure}

In fig.~(\ref{R1sofb}) I show the suppression factor of 1S states for the spherical nuclei collisions obtained by comparing the in-medium production to the production obtained  with the vacuum potential. In the left (right) panel I show the production rate for the $\Upsilon$ 1S  ($J/\psi$) for the two model potentials as a function of the impact parameter.  As inferred from the fixed $\mu_D$ studies, the  complex potential is much more effective in suppressing both quarkonia states while the real potential leads to a modest suppression in both cases.  As expected, since $J/\psi$ states
melt at a lower temperature than the $\Upsilon$ ground state, the suppression is larger for the former than from the latter. For the particular parameter set in this study, the in-medium potential supports bound 1S $\Upsilon$ and $J/\psi$ states in most of the evolution and is only able to dissolve the 
$J/\psi$ for the hottest part of the most central collisions.  However, and in spite of the different dissociation temperatures, the nuclear absorption
is approximately constant until similar values of the impact parameter, $b\sim 8$ fm. This fact is a consequence of the long formation time of the
in-medium bound states, since increasing $b$ leads to a shorter lifetime of the fireball. For the collision model I have considered, $b=8$ fm 
corresponds to the the largest impact parameter for which the fireball lifetime is longer than the quarkonia formation time.

\begin{figure}[tbp]
\includegraphics[width=.45\textwidth]{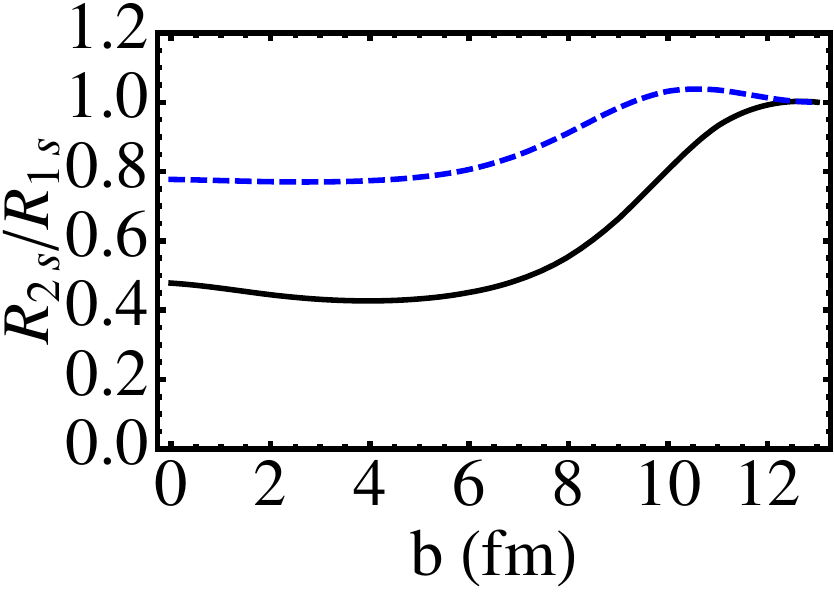}
\hfill
\includegraphics[width=.45\textwidth]{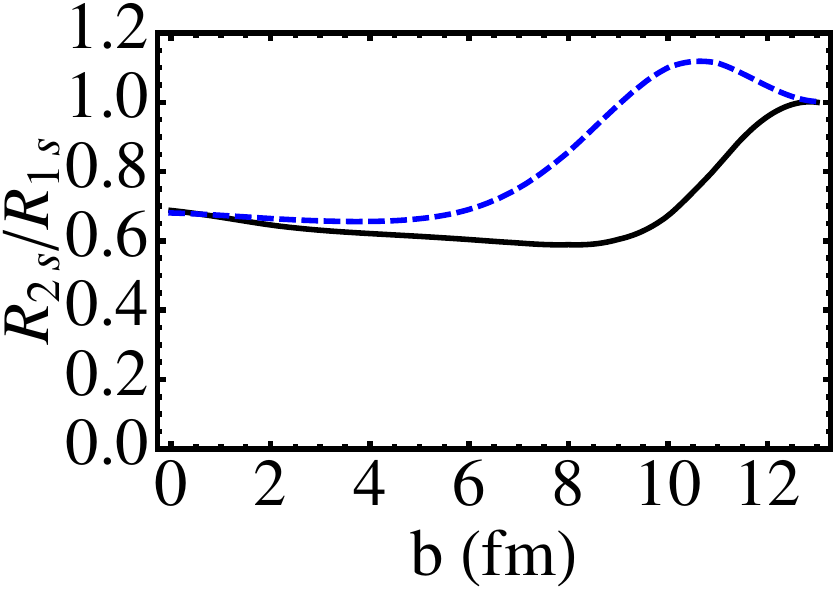}
\caption{Double ratio of the suppression factors of the 2S to the 1S state  for the spherical nuclei collisions as a function of impact parameter
$b$ for both the real  (dashed) and complex (solid) potentials. The left and right panels correspond to $\Upsilon$ and
$J/\psi$ suppressions respectively.} 
\label{R2s21sofb}
\end{figure}

In fig..~(\ref{R2s21sofb}) I show the double ratio of the suppression factors of the 2S to the 1S levels for the $\Upsilon$ and $J/\Psi$ families
both for real and complex potentials.  As in all the previous cases, the complex potential is more effective in suppressing 2S quarkonia states. 
As for the ground state case, the $b$ dependence of the suppression is relatively featureless.  For $b<6$ fm, the double ratio decreases 
slightly as $b$ increases. This is a consequence of the hotter fireball in more central collisions and the observation made in section~(\ref{brick})
that as the in-medium ground states moves towards threshold, there is an enhancement in the relative production of 2S to 1S states. However,
contrary to the fixed temperature cases in 
section~(\ref{brick}), this double ratio does not significantly cross unity for any $b$. As in the expanding case, this is a consequence of the particular choice
of parameters in this articles, since for most of the evolution the potentials I have considered support a bound level. As already mentioned, this feature is
in contrast with lattice studies of charm current-current correlators which indicate the dissolution of $J/\psi$ states soon after $T_c$. 
Finally, as for the ground state case, the double ratio rises for $b\sim 8$ fm, showing once again that this non-trivial behavior is due to the lifetime of the fireball as compared to the meson formation time.

\section{Discussion}
\label{conclusions}

Following the potential model approach to in-medium heavy quarkonia physics, the main  assumption of this work is that all  relevant interactions of 
a heavy quark pair with the QGP can be recast in a temperature dependent potential. In particular, I have assumed that all level dissociation effects 
by plasma interactions with the pair can be encoded in the potential without demanding additional processes.  If the medium is described by a real potential, the interactions lead to a reorganization of levels which are conserved in the medium. On the contrary, complex potentials naturally include dissociation processes, such as the Landau damping (see \cite{Rapp:2008tf} for a discussion on the different dissociation process),  in the imaginary par of the potential. 
 With  this approach I have calculated the  suppression pattern of different quarkonia states by a finite lifetime medium interacting
with a heavy $Q-\bar Q$ pair produced by some hard process in a singlet state. 

Potential models have the advantage that, once the correct
singlet potential  is obtained, they can be directly used to describe both the late time soft matrix element needed in quarkonia production 
as well as the current-current correlators extracted by  lattice QCD calculations. In addition, since this approach is able to treat simultaneously all  the states of the $Q-\bar Q$ system, it allows for a complete determination of the yield suppression taking into account the overlap between different vacuum and in-medium wave functions. Unfortunately,  there is currently no consensus on the functional form
of the potential which best describes lattice data in the vicinity of $T_c$. As a consequence, in this article I have used two simple potentials which lead to the sequential disappearance  of the different bound states, without any attempt to adjust to the lattice data. This oversimplified approach is sufficient to draw the main conclusions of this study and I leave a more refined treatment of the potential for future work. 

One of the main observations of this work is that, while the productions of 1S levels is dominated by the properties of the in-medium 
ground state, the production of higher excited states, such as the 2S ones, is more complicated. As I have shown in section~\ref{brick} 
these states are also influenced by the in-medium ground state in addition to the  in-medium level with the same quantum numbers. Thus, to properly describe the
suppression of quarkonia excited states it is not enough to understand the in-medium properties of these states but also the mixing of states which occurs in the medium,  since vacuum and in-medium wave functions have  non-trivial overlap.

This interplay among different states for the production of 2S states  also leads to some unexpected features in their suppression pattern. In the fixed temperature calculations in section~\ref{brick} I have shown that if the in-medium ground state is totally dissolved or if it is sufficiently close to threshold there is a strong enhancement in the relative production of 2S to 1S states. This enhancement is generic, independent of the particular details of the in-medium potential, since it is only a consequence of the 
fact that the lesser bound the in-medium levels are the wider they become, increasing their overlap with vacuum excited states. While the potential approach is more suited for bottomonium states,  this feature of the potential model calculation  goes along the trends observed in preliminary $\psi$(1S) production data  by CMS \cite{HPtalk}, which indicate a strong enhancement of the double ratio of 
$\psi(1S)$ to $J/\psi$  $R_{AA}$ at moderate $p_T$. However, a more detailed study is needed before a direct comparison with data can be made.

I have also studied the effect of an expanding medium in quarkonia production. Under the assumption that the  expansion rate is sufficiently
slow as compared to the internal processes in the plasma, I have used this potential model approach to address the dynamics of the system in a pre-asymptotic stage (see \cite{Borghini:2011ms,Dutta:2012nw} for related studies) and  I have identified the natural scale for the formation of in-medium states, the ground state period.  While for this 
exploratory study  I have used low binding energies for the different vacuum quarkonia states, these formation time effects are important
and make the quarkonia production mostly insensitive to the hottest part of the medium evolution.  This observation suggests that the suppression pattern of quarkonia in nuclear collisions must have a modest dependence on the colliding energy since the higher temperatures reached  in energetic collisions are rapidly relaxed due to the strong longitudinal expansion. Remarkably, this is precisely the trend observed in comparing suppression data from SPS to the LHC  \cite{Gonin:1996wn,Abreu:1997jh,Adare:2006ns,Aad:2010aa,Abelev:2012rv,Chatrchyan:2012np}. However, a much more detailed analysis than the one in this work must be performed before we can determine the relevance of these effects for the observed patterns in suppression data. 

These formation time effects have also a important effect in the pattern of suppression of different quarkonia levels. While for a static medium there is an enhancement of the relative abundance of excited to ground states whenever the in-medium ground state is dissolved or close to threshold, the situation in an expanding medium is more complicated. 
In fact, even if at the early stages of
the system evolution the medium is sufficiently hot to dissolve the ground state of a given quarkonia family, if for most of the evolution the ground state is present,  the enhancement
in the double ratio of 2S to 1S states observed in the static medium will not have time to develop. Once again, this observation has profound consequences for quarkonia physics at the LHC:
If the anomalous suppression pattern observed in preliminary CMS data can be attributed to this mixing effect between states below and above the transitions, 
the relative enhancement of $\psi(2S)$ production indicates that $J/\psi$ mets for the most part of the hot medium evolution at the LHC, which, in turn, points towards a low melting temperature of in-medium quarkonia

In summary, the potential model calculation I have explored in this paper is a simple and powerful method to describe the suppression patter of quarkonia  in heavy ion collisions.  
The analysis in this work has been oversimplified not only because of the simplified potentials that I have used, but also for the simplified treatment
of quarkonia production which, in particular, neglects octect contribution, the lack of energy loss of the pair, the interaction of more than one pair created in the collision, initial state effects, etc. All these processes must be address before drawing any firm conclusion about quarkonia suppression in data. Nevertheless, some of the features I have discussed, such as the complicated interplay among vacuum and medium levels are completely generic of this approach, independently of the details of the potential.

\acknowledgments

I thank F. Arleo, E. G. Ferreiro, A. Palacios and  J. Soto for useful discussions. I am   supported by a Ram\'on y Cajal fellowship. I also  acknowledge financial support by the research grants FPA2010-20807, 2009SGR502 and by the Consolider CPAN project.



\bibliography{bib}

\providecommand{\href}[2]{#2}\begingroup\raggedright\begin{thebibliography}{10}

\bibitem{Adcox:2004mh}
{\bf PHENIX Collaboration} Collaboration, K.~Adcox et~al., {\it {Formation of
  dense partonic matter in relativistic nucleus-nucleus collisions at RHIC:
  Experimental evaluation by the PHENIX collaboration}},  {\em Nucl.Phys.} {\bf
  A757} (2005) 184--283, [\href{http://xxx.lanl.gov/abs/nucl-ex/0410003}{{\tt
  nucl-ex/0410003}}].

\bibitem{Adams:2005dq}
{\bf STAR Collaboration} Collaboration, J.~Adams et~al., {\it {Experimental and
  theoretical challenges in the search for the quark gluon plasma: The STAR
  Collaboration's critical assessment of the evidence from RHIC collisions}},
  {\em Nucl.Phys.} {\bf A757} (2005) 102--183,
  [\href{http://xxx.lanl.gov/abs/nucl-ex/0501009}{{\tt nucl-ex/0501009}}].

\bibitem{Muller:2012zq}
B.~Muller, J.~Schukraft, and B.~Wyslouch, {\it {First Results from Pb+Pb
  collisions at the LHC}},  \href{http://xxx.lanl.gov/abs/1202.3233}{{\tt
  arXiv:1202.3233}}.

\bibitem{Matsui:1986dk}
T.~Matsui and H.~Satz, {\it {J/psi Suppression by Quark-Gluon Plasma
  Formation}},  {\em Phys.Lett.} {\bf B178} (1986) 416.

\bibitem{Lansberg:2006dh}
J.~Lansberg, {\it {$J/\psi$, $\psi$ ' and $\upsilon$ production at hadron
  colliders: A Review}},  {\em Int.J.Mod.Phys.} {\bf A21} (2006) 3857--3916,
  [\href{http://xxx.lanl.gov/abs/hep-ph/0602091}{{\tt hep-ph/0602091}}].

\bibitem{Rapp:2008tf}
R.~Rapp, D.~Blaschke, and P.~Crochet, {\it {Charmonium and bottomonium
  production in heavy-ion collisions}},  {\em Prog.Part.Nucl.Phys.} {\bf 65}
  (2010) 209--266, [\href{http://xxx.lanl.gov/abs/0807.2470}{{\tt
  arXiv:0807.2470}}].

\bibitem{BraunMunzinger:2000px}
P.~Braun-Munzinger and J.~Stachel, {\it {(Non)thermal aspects of charmonium
  production and a new look at J / psi suppression}},  {\em Phys.Lett.} {\bf
  B490} (2000) 196--202, [\href{http://xxx.lanl.gov/abs/nucl-th/0007059}{{\tt
  nucl-th/0007059}}].

\bibitem{Thews:2000rj}
R.~L. Thews, M.~Schroedter, and J.~Rafelski, {\it {Enhanced J / psi production
  in deconfined quark matter}},  {\em Phys.Rev.} {\bf C63} (2001) 054905,
  [\href{http://xxx.lanl.gov/abs/hep-ph/0007323}{{\tt hep-ph/0007323}}].

\bibitem{Gorenstein:2000ck}
M.~I. Gorenstein, A.~Kostyuk, H.~Stoecker, and W.~Greiner, {\it {Statistical
  coalescence model with exact charm conservation}},  {\em Phys.Lett.} {\bf
  B509} (2001) 277--282, [\href{http://xxx.lanl.gov/abs/hep-ph/0010148}{{\tt
  hep-ph/0010148}}].

\bibitem{Grandchamp:2001pf}
L.~Grandchamp and R.~Rapp, {\it {Thermal versus direct J / Psi production in
  ultrarelativistic heavy ion collisions}},  {\em Phys.Lett.} {\bf B523} (2001)
  60--66, [\href{http://xxx.lanl.gov/abs/hep-ph/0103124}{{\tt
  hep-ph/0103124}}].

\bibitem{Gonin:1996wn}
{\bf NA50 Collaboration} Collaboration, M.~Gonin et~al., {\it {Anomalous J /
  psi suppression in Pb + Pb collisions at 158-A-GeV/c}},  {\em Nucl.Phys.}
  {\bf A610} (1996) 404C--417C.

\bibitem{Abreu:1997jh}
{\bf NA50 Collaboration} Collaboration, M.~Abreu et~al., {\it {Anomalous J /
  psi suppression in Pb - Pb interactions at 158 GeV/c per nucleon}},  {\em
  Phys.Lett.} {\bf B410} (1997) 337--343.

\bibitem{Adare:2006ns}
{\bf PHENIX Collaboration} Collaboration, A.~Adare et~al., {\it {J/psi
  Production vs Centrality, Transverse Momentum, and Rapidity in Au+Au
  Collisions at s(NN)**(1/2) = 200-GeV}},  {\em Phys.Rev.Lett.} {\bf 98} (2007)
  232301, [\href{http://xxx.lanl.gov/abs/nucl-ex/0611020}{{\tt
  nucl-ex/0611020}}].

\bibitem{Aad:2010aa}
{\bf Atlas Collaboration} Collaboration, G.~Aad et~al., {\it {Measurement of
  the centrality dependence of $J/{\psi}$ yields and observation of Z
  production in lead-lead collisions with the ATLAS detector at the LHC}},
  {\em Phys.Lett.} {\bf B697} (2011) 294--312,
  [\href{http://xxx.lanl.gov/abs/1012.5419}{{\tt arXiv:1012.5419}}].

\bibitem{Abelev:2012rv}
{\bf ALICE Collaboration} Collaboration, B.~Abelev et~al., {\it {J/psi
  production at low transverse momentum in Pb-Pb collisions at sqrt(sNN) = 2.76
  TeV}},  \href{http://xxx.lanl.gov/abs/1202.1383}{{\tt arXiv:1202.1383}}.

\bibitem{Chatrchyan:2012np}
{\bf CMS Collaboration} Collaboration, S.~Chatrchyan et~al., {\it {Suppression
  of non-prompt J/psi, prompt J/psi, and Y(1S) in PbPb collisions at sqrt(sNN)
  = 2.76 TeV}},  {\em JHEP} {\bf 1205} (2012) 063,
  [\href{http://xxx.lanl.gov/abs/1201.5069}{{\tt arXiv:1201.5069}}].

\bibitem{HPtalk}
{\bf CMS Collaboration} Collaboration, T.~Dhams, {\it talk given at the {\it
  5th international conference on hard and electromagnetic probes of
  high-energy nuclear collisions} (hp2012).}, .

\bibitem{Reed:2011fr}
R.~Reed, {\it {Measuring the Upsilon Nuclear Modification Factor at STAR}},
  {\em J.Phys.G} {\bf G38} (2011) 124185,
  [\href{http://xxx.lanl.gov/abs/1109.3891}{{\tt arXiv:1109.3891}}].

\bibitem{Asakawa:2003re}
M.~Asakawa and T.~Hatsuda, {\it {J / psi and eta(c) in the deconfined plasma
  from lattice QCD}},  {\em Phys.Rev.Lett.} {\bf 92} (2004) 012001,
  [\href{http://xxx.lanl.gov/abs/hep-lat/0308034}{{\tt hep-lat/0308034}}].

\bibitem{Datta:2003ww}
S.~Datta, F.~Karsch, P.~Petreczky, and I.~Wetzorke, {\it {Behavior of
  charmonium systems after deconfinement}},  {\em Phys.Rev.} {\bf D69} (2004)
  094507, [\href{http://xxx.lanl.gov/abs/hep-lat/0312037}{{\tt
  hep-lat/0312037}}].

\bibitem{Morrin:2005zq}
R.~Morrin, A.~O~Cais, M.~Oktay, M.~Peardon, J.~Skullerud, et~al., {\it
  {Charmonium spectral functions in $N_f=2$ QCD}},  {\em PoS} {\bf LAT2005}
  (2006) 176, [\href{http://xxx.lanl.gov/abs/hep-lat/0509115}{{\tt
  hep-lat/0509115}}].

\bibitem{Jakovac:2006sf}
A.~Jakovac, P.~Petreczky, K.~Petrov, and A.~Velytsky, {\it {Quarkonium
  correlators and spectral functions at zero and finite temperature}},  {\em
  Phys.Rev.} {\bf D75} (2007) 014506,
  [\href{http://xxx.lanl.gov/abs/hep-lat/0611017}{{\tt hep-lat/0611017}}].

\bibitem{Aarts:2007pk}
G.~Aarts, C.~Allton, M.~B. Oktay, M.~Peardon, and J.-I. Skullerud, {\it
  {Charmonium at high temperature in two-flavor QCD}},  {\em Phys.Rev.} {\bf
  D76} (2007) 094513, [\href{http://xxx.lanl.gov/abs/0705.2198}{{\tt
  arXiv:0705.2198}}].

\bibitem{Aarts:2011sm}
G.~Aarts, C.~Allton, S.~Kim, M.~Lombardo, M.~Oktay, et~al., {\it {What happens
  to the Upsilon and $eta_b$ in the quark-gluon plasma? Bottomonium spectral
  functions from lattice QCD}},  {\em JHEP} {\bf 1111} (2011) 103,
  [\href{http://xxx.lanl.gov/abs/1109.4496}{{\tt arXiv:1109.4496}}].

\bibitem{Ding:2012sp}
H.-T. Ding, A.~Francis, O.~Kaczmarek, F.~Karsch, H.~Satz, et~al., {\it
  {Charmonium properties in hot quenched lattice QCD}},
  \href{http://xxx.lanl.gov/abs/1204.4945}{{\tt arXiv:1204.4945}}.

\bibitem{Karsch:1987pv}
F.~Karsch, M.~Mehr, and H.~Satz, {\it {Color Screening and Deconfinement for
  Bound States of Heavy Quarks}},  {\em Z.Phys.} {\bf C37} (1988) 617.

\bibitem{Digal:2001ue}
S.~Digal, P.~Petreczky, and H.~Satz, {\it {Quarkonium feed down and sequential
  suppression}},  {\em Phys.Rev.} {\bf D64} (2001) 094015,
  [\href{http://xxx.lanl.gov/abs/hep-ph/0106017}{{\tt hep-ph/0106017}}].

\bibitem{Shuryak:2004tx}
E.~V. Shuryak and I.~Zahed, {\it {Towards a theory of binary bound states in
  the quark gluon plasma}},  {\em Phys.Rev.} {\bf D70} (2004) 054507,
  [\href{http://xxx.lanl.gov/abs/hep-ph/0403127}{{\tt hep-ph/0403127}}].

\bibitem{Wong:2004zr}
C.-Y. Wong, {\it {Heavy quarkonia in quark-gluon plasma}},  {\em Phys.Rev.}
  {\bf C72} (2005) 034906, [\href{http://xxx.lanl.gov/abs/hep-ph/0408020}{{\tt
  hep-ph/0408020}}].

\bibitem{Alberico:2005xw}
W.~Alberico, A.~Beraudo, A.~De~Pace, and A.~Molinari, {\it {Heavy quark bound
  states above T(c)}},  {\em Phys.Rev.} {\bf D72} (2005) 114011,
  [\href{http://xxx.lanl.gov/abs/hep-ph/0507084}{{\tt hep-ph/0507084}}].

\bibitem{Mannarelli:2005pz}
M.~Mannarelli and R.~Rapp, {\it {Hadronic modes and quark properties in the
  quark-gluon plasma}},  {\em Phys.Rev.} {\bf C72} (2005) 064905,
  [\href{http://xxx.lanl.gov/abs/hep-ph/0505080}{{\tt hep-ph/0505080}}].

\bibitem{Mocsy:2005qw}
A.~Mocsy and P.~Petreczky, {\it {Quarkonia correlators above deconfinement}},
  {\em Phys.Rev.} {\bf D73} (2006) 074007,
  [\href{http://xxx.lanl.gov/abs/hep-ph/0512156}{{\tt hep-ph/0512156}}].

\bibitem{Mocsy:2007yj}
A.~Mocsy and P.~Petreczky, {\it {Can quarkonia survive deconfinement?}},  {\em
  Phys.Rev.} {\bf D77} (2008) 014501,
  [\href{http://xxx.lanl.gov/abs/0705.2559}{{\tt arXiv:0705.2559}}].

\bibitem{Mocsy:2007jz}
A.~Mocsy and P.~Petreczky, {\it {Color screening melts quarkonium}},  {\em
  Phys.Rev.Lett.} {\bf 99} (2007) 211602,
  [\href{http://xxx.lanl.gov/abs/0706.2183}{{\tt arXiv:0706.2183}}].

\bibitem{Cabrera:2006wh}
D.~Cabrera and R.~Rapp, {\it {T-Matrix Approach to Quarkonium Correlation
  Functions in the QGP}},  {\em Phys.Rev.} {\bf D76} (2007) 114506,
  [\href{http://xxx.lanl.gov/abs/hep-ph/0611134}{{\tt hep-ph/0611134}}].

\bibitem{Laine:2007gj}
M.~Laine, {\it {A Resummed perturbative estimate for the quarkonium spectral
  function in hot QCD}},  {\em JHEP} {\bf 0705} (2007) 028,
  [\href{http://xxx.lanl.gov/abs/0704.1720}{{\tt arXiv:0704.1720}}].

\bibitem{Kaczmarek:2005ui}
O.~Kaczmarek and F.~Zantow, {\it {Static quark anti-quark interactions in zero
  and finite temperature QCD. I. Heavy quark free energies, running coupling
  and quarkonium binding}},  {\em Phys.Rev.} {\bf D71} (2005) 114510,
  [\href{http://xxx.lanl.gov/abs/hep-lat/0503017}{{\tt hep-lat/0503017}}].

\bibitem{Eichten:1979ms}
E.~Eichten, K.~Gottfried, T.~Kinoshita, K.~Lane, and T.-M. Yan, {\it
  {Charmonium: Comparison with Experiment}},  {\em Phys.Rev.} {\bf D21} (1980)
  203.

\bibitem{Laine:2006ns}
M.~Laine, O.~Philipsen, P.~Romatschke, and M.~Tassler, {\it {Real-time static
  potential in hot QCD}},  {\em JHEP} {\bf 0703} (2007) 054,
  [\href{http://xxx.lanl.gov/abs/hep-ph/0611300}{{\tt hep-ph/0611300}}].

\bibitem{Beraudo:2007ky}
A.~Beraudo, J.-P. Blaizot, and C.~Ratti, {\it {Real and imaginary-time Q anti-Q
  correlators in a thermal medium}},  {\em Nucl.Phys.} {\bf A806} (2008)
  312--338, [\href{http://xxx.lanl.gov/abs/0712.4394}{{\tt arXiv:0712.4394}}].

\bibitem{Escobedo:2008sy}
M.~A. Escobedo and J.~Soto, {\it {Non-relativistic bound states at finite
  temperature (I): The Hydrogen atom}},  {\em Phys.Rev.} {\bf A78} (2008)
  032520, [\href{http://xxx.lanl.gov/abs/0804.0691}{{\tt arXiv:0804.0691}}].

\bibitem{Brambilla:2008cx}
N.~Brambilla, J.~Ghiglieri, A.~Vairo, and P.~Petreczky, {\it {Static
  quark-antiquark pairs at finite temperature}},  {\em Phys.Rev.} {\bf D78}
  (2008) 014017, [\href{http://xxx.lanl.gov/abs/0804.0993}{{\tt
  arXiv:0804.0993}}].

\bibitem{Brambilla:2010vq}
N.~Brambilla, M.~A. Escobedo, J.~Ghiglieri, J.~Soto, and A.~Vairo, {\it {Heavy
  Quarkonium in a weakly-coupled quark-gluon plasma below the melting
  temperature}},  {\em JHEP} {\bf 1009} (2010) 038,
  [\href{http://xxx.lanl.gov/abs/1007.4156}{{\tt arXiv:1007.4156}}].

\bibitem{Sharma:2009hn}
R.~Sharma, I.~Vitev, and B.-W. Zhang, {\it {Light-cone wave function approach
  to open heavy flavor dynamics in QCD matter}},  {\em Phys.Rev.} {\bf C80}
  (2009) 054902, [\href{http://xxx.lanl.gov/abs/0904.0032}{{\tt
  arXiv:0904.0032}}].

\bibitem{Strickland:2011mw}
M.~Strickland, {\it {Thermal Upsilon(1s) and $chi_b1$ suppression in
  sqrt(sNN)=2.76 TeV Pb-Pb collisions at the LHC}},  {\em Phys.Rev.Lett.} {\bf
  107} (2011) 132301, [\href{http://xxx.lanl.gov/abs/1106.2571}{{\tt
  arXiv:1106.2571}}].

\bibitem{Emerick:2011xu}
A.~Emerick, X.~Zhao, and R.~Rapp, {\it {Bottomonia in the Quark-Gluon Plasma
  and their Production at RHIC and LHC}},  {\em Eur.Phys.J.} {\bf A48} (2012)
  72, [\href{http://xxx.lanl.gov/abs/1111.6537}{{\tt arXiv:1111.6537}}].

\bibitem{Song:2011nu}
T.~Song, K.~C. Han, and C.~M. Ko, {\it {Bottomonia suppression in heavy-ion
  collisions}},  {\em Phys.Rev.} {\bf C85} (2012) 014902,
  [\href{http://xxx.lanl.gov/abs/1109.6691}{{\tt arXiv:1109.6691}}].

\bibitem{Bodwin:1994jh}
G.~T. Bodwin, E.~Braaten, and G.~P. Lepage, {\it {Rigorous QCD analysis of
  inclusive annihilation and production of heavy quarkonium}},  {\em Phys.Rev.}
  {\bf D51} (1995) 1125--1171,
  [\href{http://xxx.lanl.gov/abs/hep-ph/9407339}{{\tt hep-ph/9407339}}].

\bibitem{Borghini:2011ms}
N.~Borghini and C.~Gombeaud, {\it {Heavy quarkonia in a medium as a quantum
  dissipative system: Master equation approach}},  {\em Eur.Phys.J.} {\bf C72}
  (2012) 2000, [\href{http://xxx.lanl.gov/abs/1109.4271}{{\tt
  arXiv:1109.4271}}].

\bibitem{Dutta:2012nw}
N.~Dutta and N.~Borghini, {\it {Sequential suppression of quarkonia and
  high-energy nucleus-nucleus collisions}},
  \href{http://xxx.lanl.gov/abs/1206.2149}{{\tt arXiv:1206.2149}}.

\end{thebibliography}\endgroup
\bibliographystyle{JHEP}


\end{document}